\documentclass[cameraready]{Interspeech}
\usepackage{booktabs}
\usepackage{multirow, graphicx}
\usepackage{times}            %
\usepackage[english]{babel}   %
\usepackage[ansinew]{inputenc}%
\usepackage[T1]{fontenc}      %
\usepackage{amsmath,amssymb}
\usepackage{graphicx}
\usepackage[capitalize,noabbrev]{cleveref}

\newcommand\norm[1]{\lVert#1\rVert}

\usepackage{mathtools}
\usepackage{xfrac}

\newcommand{\D}[1]{\mathrm{d}{#1}}

\newcommand{\ts}{s_\mathrm{time}}
\newcommand{\ty}{y_\mathrm{time}}

\newcommand{\nfft}{n_\mathrm{{fft}}}
\newcommand{\nhop}{n_\mathrm{{hop}}}

\newcommand{\lti}{\lambda_{t_i}}
\newcommand{\ltimo}{\lambda_{t_{i-1}}} %
\newcommand{\ti}{t_i}
\newcommand{\timo}{t_{i-1}}

\newcommand{\Tmax}{T_{\text{max}}}

\newcommand{\scorem}{s_{\theta}}
\newcommand{\score}{\nabla_{  x_t} \log p_t(  x_t| y)}
\newcommand{\epsm}{\rho_{\theta}}

\newcommand{\rtol}{r_{\mathrm{tol}}}
\newcommand{\atol}{a_{\mathrm{tol}}}

\usepackage{comment}

\newcommand{\Ldsm}{L_{\text{DSM}}}

\newcommand{\smin}{\mathrm{\sigma_{\text{min}}}}
\newcommand{\smax}{\mathrm{\sigma_{\text{max}}}}

\usepackage[font=small,skip=0pt]{caption}
\usepackage[nolist, nohyperlinks]{acronym}
\usepackage{algorithm}
\usepackage{algpseudocode}

\begin{acronym}
\acro{bb}[BB]{Brownian bridge}
\acro{rk}[RK]{Runge-Kutta}
\acro{erk}[expRK]{exponential Runge-Kutta}

\acro{SQA}{speech quality assessment}
\acro{eum}[EuM]{Euler-Maruyama}
\acro{mos}[MOS]{Mean Opinion Score}
\acro{sr}[SR]{Speech Restoration}
\acro{ouve}[OUVE]{Ornstein-Uhlenbeck Variance Exploding}
\acro{fouve}[fOUVE]{fixed Ornstein-Uhlenbeck Variance Exploding}
\acro{bbed}[BBED]{Brownian-Bridge with Exploding Diffusion}
\acro{ot}[OT]{Optimal Transport}
\acro{ode}[ODE]{Ordinary Differential Equation}
\acro{nn}[NN]{Neural Network}
\acro{sgm}[SGM]{score-based generative model}
\acro{snr}[SNR]{signal-to-noise ratio}
\acro{gan}[GAN]{generative adversarial network}
\acro{vae}[VAE]{variational autoencoder}
\acro{dpm}[dpm]{denoising diffusion probabilistic model}
\acro{stft}[STFT]{short-time Fourier transform}
\acro{istft}[iSTFT]{inverse short-time Fourier transform}
\acro{rsde}[RSDE]{reverse stochastic differential equation}
\acro{pfode}[PF-ODE]{probability flow ordinary differential equation}
\acro{ald}[ALD]{anneleaed Langevin Dynamics}

\acro{dpm}[DPM]{Diffusion Probabilistic Model}
\acro{sde}[SDE]{stochastic differential equation}
\acro{ode}[ODE]{ordinary differential equation}
\acro{ou}[OU]{Ornstein-Uhlenbeck}
\acro{pc}[PC]{Predictor-Corrector}
\acro{ve}[VE]{Variance Exploding}
\acro{dnn}[DNN]{deep neural network}
\acro{pesq}[PESQ]{Perceptual Evaluation of Speech Quality}
\acro{bwe}[BWE]{bandwidth extension}
\acro{rir}[RIR]{room impulse response}

\acro{tf}[T-F]{time-frequency}
\acro{elbo}[ELBO]{evidence lower bound}
\acro{WPE}{weighted prediction error}
\acro{PSD}{power spectral density}
\acro{RIR}{room impulse response}
\acro{SNR}{signal-to-noise ratio}
\acro{dpmsol}[DPM-pS]{DPM-Solver-p}
\acro{isdesol}[iSDE-$p$S-$\kappa$]{iSDE-Solver-$p$-$\kappa$}

\acro{LSTM}{long short-term memory}
\acro{POLQA}{Perceptual Objectve Listening Quality Analysis}
\acro{SDR}{signal-to-distortion ratio}
\acro{ESTOI}{Extended Short-Term Objective Intelligibility}
\acro{ELR}{early-to-late reverberation ratio}
\acro{TCN}{temporal convolutional network}
\acro{DRR}{direct-to-reverberant ratio}
\acro{nfe}[NFE]{number of function evaluation}
\acro{rtf}[RTF]{real-time factor}
\acro{em}[EuM]{Euler-Maruyama}
\acro{isde}[iSDE]{interpolation SDE}
\acro{ald}[ALD]{annealed Langevin Dynamics}
\acro{crp}[CRP]{correcting the reverse process}
\acro{dsm}[DSM]{denoising score matching}
\end{acronym}

\title{A Fast Solver for Interpolating Stochastic Differential Equation Diffusion Models for Speech Restoration}

\author[affiliation={1,2}, orcid=0000-0002-0847-7896]{Bunlong}{Lay}
\author[affiliation={1}, orcid=0000-0002-8678-4699]{Timo}{Gerkmann}
\address{1~Universität Hamburg, Germany}

\address{
    $^1$ University of Hamburg, Signal Processing (SP), Bundesstr. 56b, Hamburg, Germany \\
    $^2$ HITeC-Hamburg, Bundesstr. 56b, Hamburg, Germany 
}

\email{bunlong.lay@uni-hamburg.de, timo.gerkmann@uni-hamburg.de}


\usepackage{comment}

\begin{document}
\maketitle
\begin{abstract}
\acp{dpm} are a well-established class of diffusion models for unconditional image generation, while SGMSE+ is a well-established conditional diffusion model for speech enhancement. One of the downsides of diffusion models is that solving the reverse process requires many evaluations of a large Neural Network. Although advanced fast sampling solvers have been developed for \acp{dpm}, they are not directly applicable to models such as SGMSE+ due to differences in their diffusion processes. Specifically, \acp{dpm} transform between the data distribution and a standard Gaussian distribution, whereas SGMSE+ interpolates between the target distribution and a noisy observation. This work first develops a formalism of interpolating Stochastic Differential Equations (iSDEs) that includes SGMSE+, and second proposes a solver for iSDEs. The proposed solver enables fast sampling with as few as 10 Neural Network evaluations across multiple speech restoration tasks.\footnote{Published code \url{https://github.com/sp-uhh/fast_solver_interpolating_sde}}
\end{abstract}

\section{Introduction}
The goal of \ac{sr} is to retrieve the clean speech signal from a degraded signal that has been affected by some corruption. 
Traditional methods attempt to leverage the statistical relationships between the clean speech signal and the degraded signal. Various machine learning techniques have been suggested, treating \ac{sr} as a predictive learning task \cite{wang2018supervised, luo2019conv}.

Diverging from predictive approaches, which establish a direct mapping from degraded to clean speech, generative approaches focus on learning a prior distribution over clean speech data. Recently, a category of generative models known as \emph{diffusion models} (or \emph{score-based generative models}) has been introduced to the realm of \ac{sr} \cite{lu2021study, lu2022conditional, welker2022speech, richter_sgmse}. In \cite{lu2021study, lu2022conditional}, the concept involves gradually adding Gaussian noise to the data through a discrete and fixed Markov chain, referred to as the \emph{forward process}, thereby transforming the data into a tractable distribution like a Gaussian distribution. Subsequently, a \ac{nn} is trained to reverse this diffusion process in a so-called \emph{reverse process} \cite{ho2020denoising}. As the step size between two discrete Markov chain states approaches zero, the discrete Markov chain transforms into a continuous-time \ac{sde} under mild constraints. The use of \acp{sde} provides greater flexibility and opportunities compared to methods based on discrete Markov chains \cite{song2021sde}. For \ac{sr}, a \ac{sde} has been originally proposed in \cite{welker2022speech, richter_sgmse}. Notably, \acp{sde} enable the application of general-purpose \ac{sde} solvers for numerically integrating the reverse process, thereby influencing performance and the number of iteration steps. An \ac{sde} can be viewed as resulting in a transformation between two specified distributions, with one designated as the initial distribution and the other as the terminating distribution of the \ac{sde}. In the context of SGMSE+ \cite{richter_sgmse}, recently different SDEs \cite{richter_sgmse, lay202interspeech} have been introduced, with the initial distribution being the clean speech data and the terminating distribution being centered around the noisy mixture. Hence, these SDEs result in a stochastic interpolation between the clean speech signal and the noisy mixture. Within mild constraints, it is possible to identify a reverse \ac{ode}, known as the \ac{pfode} \cite{song2019generative}, and different reverse \acp{sde} \cite{Zhang2022FastSO} for each forward \ac{sde} that effectively reverses the forward process. This reverse \ac{ode}/\ac{sde} starts from the degraded signal plus some Gaussian noise and ends at an estimate of the clean speech. The key difference between a reverse SDE and a reverse ODE is that the reverse SDE adds Gaussian noise during the reverse process, which enables the sampler to explore different regions of the model's learned distribution, rather than collapsing to one deterministic trajectory. As this process assumes that the degraded signal is available, we call this formulation \emph{conditional diffusion process}. In contrast, in \emph{unconditional diffusion process}, the terminating distribution is an uninformed standard Gaussian and thus independent of the degraded signal. Solving the reverse \ac{pfode} generates a conditional estimate of the clean speech signal.

Since the success of the SGMSE+'s \cite{welker2022speech, richter_sgmse} conditional diffusion process formulation \cite{lu2022conditional}, many other \acp{isde} \cite{lay202interspeech, vpidm, trachu24_interspeech, qiu2023sebridge, jun2025snraligned, lee2025flowse} have been applied for score-based \ac{sr}. Even some Schr\"odinger Bridges \cite{jukic2024sbse, Han2025FewstepAS, richter2024investigating} can be formulated as an \acp{isde}. However, no general mathematical formulation has been developed to unify these \acp{isde}. In this work, we develop a formalism for arbitrary interpolating \acp{sde} from which all \acp{isde} can be derived. Under this formalism, the unconditional generation task becomes a special case, where the mean interpolates towards 0, instead of towards a degraded signal as proposed in SGMSE+. For unconditional diffusion processes, many fast sampling methods \cite{Lu2022DPMSolver, Lu2022DPMSolverFS, zheng2023dpmsolverv, Zhang2022FastSO} have been developed, where \cite{Lu2022DPMSolver} is one of the first prominent fast solvers, known as the \emph{DPM-Solver} or \emph{DPM-pS}. All fast sampling methods achieved better quality with fewer \acp{nfe}, where \acp{nfe} denotes the number of times a \ac{nn} is called when solving the reverse process. However, these fast solvers can not be directly applied to \acp{isde}, as the solvers were only derived for the unconditional diffusion process. Hence, in this work, we develop a novel solver, based on DPM-Solver, for the conditional diffusion process formulated by \acp{isde}. The formulation of \acp{isde} in combination with the novel solver paves the road for future work to develop other DPM-Solver variants \cite{Lu2022DPMSolverFS, zheng2023dpmsolverv} for conditional diffusion.

The DPM-Solver \cite{Lu2022DPMSolver} is a fast \ac{ode} solver that has demonstrated its superiority over other solvers, such as \ac{eum} or "classical RK45" \cite{butcher2016numerical}. An essential idea of the DPM-Solver is to adopt the so-called \ac{erk}, a well-known family of \ac{ode} solvers, to the \ac{pfode}. Different from classical \ac{rk} methods, the \ac{erk} method allows to exactly integrate the linear part of the solution, instead of only approximating it. As a result, the DPM-Solver achieves better quality with fewer reverse steps for unconditional diffusion. Inspired by \cite{Lu2022DPMSolver}, in this work, based on the \ac{erk} solvers, we derive a novel solver for conditional diffusion tasks such as SGMSE+. To this end, we develop a fast \ac{ode} sampler for solving the \ac{pfode} for a wide class of \acp{isde} and provide an analysis of this fast sampler against different samplers from the literature on various audio tasks. We experiment on \ac{sr} tasks such as Noise reduction, \ac{bwe}, Declipping, MP3 decoding, and Dereverberation. We show that the proposed solver requires only 10 \acp{nfe} for restoring the clean speech target, achieving similar performance as the higher-order adaptive RK45 solver, which has more than 40 \acp{nfe}.

\section{Diffusion Models}

\begin{table*}[t]
\vspace{-0.5cm}
\renewcommand{\arraystretch}{1.8}
    \centering
    \caption{Summary of existing \acp{isde} for audio tasks. The \ac{sde} is defined by \eqref{eq:fsde}, where only $\gamma(t)$ is needed to define the diffusion coefficient $f_t(x_t,y)$ (see \eqref{eq:interpol-drift}). The distribution of $x_t$ is given by its mean and variance-evolution $\mu_t, \sigma_t$, where the interpolation function $k(t)$ determines the mean-evolution $\mu_t$ (see \eqref{eq:mean_inter}). $c,r>0, \smin, \smax, \gamma_0>0$ are parameters of the diffusion and drift coefficient.} \label{tab:sde}
  \small
    \setlength{\tabcolsep}{2pt} %
    \begin{tabular}{c|c|c|c|c|c|c}
     Name & $\Tmax$ &  $T$&$\gamma(t)$ & $g(t)$ & $k(t)$ & $\sigma_t$ \\
     \hline
     \hline
     fOUVE & $\infty$ & 1 & $\gamma_0$ 
& $\smin(\frac{\smax}{\smin})^t\sqrt{2\ln(\frac{\smax}{\smin}) + 2\gamma_0}$ 
& $1-e^{-\gamma_0 t}$ 
& $\smin(\frac{\smax}{\smin})^{t}$  \\
     \hline
     \multirow{2}{*}{OUVE \cite{welker2022speech, richter_sgmse}} & \multirow{2}{*}{$\infty$} & \multirow{2}{*}{1} & \multirow{2}{*}{$\gamma_0$} & \multirow{2}{*}{$\smin(\frac{\smax}{\smin})^t\sqrt{2\ln(\frac{\smax}{\smin})}$} & \multirow{2}{*}{$1-e^{-\gamma_0 t}$} & $\sqrt{[(\frac{\smax}{\smin})^{2t} - e^{-2\gamma_0 t}]}K$ \\
     & & & & & & $K=\smin\sqrt{\frac{\ln(\smax/\smin)}{(\gamma_0 + \ln(\smax/\smin))}}$ \\
     \hline
      BBED \cite{lay202interspeech} & 1 & 0.999& $\frac{1}{1-t}$ &  $cr^t$ & $t$ & analytically not solvable  \\ 
     \hline
      Optimal Transport \cite{lee2025flowse} & 1 &0.999 & $\frac{1}{1-t}$ &  $\smax \sqrt{\frac{2t}{1-t}}$ & $t$ & $\smax \cdot t$ \\ 
     \hline
   Brownian Bridge \cite{trachu24_interspeech, qiu2023sebridge}  & 1 &0.999 & $\frac{1}{1-t}$ & 1 & $t$ & $t(1-t)$ \\ 
   \hline
    \end{tabular}
\end{table*}

\subsection{Stochastic Differential Equations} \label{sec:sde}
Let $s$ be clean speech and $y$ be a degraded version of $s$. 
Following the approach in \cite{richter_sgmse, lay202interspeech}, we model the forward process of the score-based generative model with an \ac{sde} defined on $0 \leq t < T_{\text{max}}$:
\begin{equation} \label{eq:fsde}
    \D{x_t} = \underbrace{f_t(x_t,y)}_{A(t)x_t + a(t)}\D{t} + g(t)\D{{ w}},
\end{equation}
where $A(t),a(t)$ are continuous functions in $t$. Since the drift coefficient $f_t$ and diffusion coefficient $g$ have only linear dependency on $x_t$, we call such an \ac{sde} a linear \ac{sde}. Furthermore, $w$ is the standard Wiener process \cite{kara_and_shreve}, $x_t$ is the current process state with initial condition $x_0 =  s$, and $t$ is a continuous diffusion time-step variable that describes the progress of the process that ends in the last diffusion time-step $T_{\text{max}}$.
The term $f_t( x_t, y) \D{t}$ can be integrated by Lebesgue integration \cite{rudin}, and $g(t)\D{{w}}$ follows It\^o integration \cite{kara_and_shreve}. 
In the considered \acp{sde}, the diffusion coefficient $g$ regulates the amount of Gaussian noise that is added to the process, and the drift coefficient $f$ affects the mean and standard deviation of $x_t$. With this model, the process state $x_t$ follows a Gaussian distribution \cite[Ch. 5]{sarkka2019sde}, called the \emph{perturbation kernel}:
\begin{equation}
\label{eq:perturbation-kernel}
    p_{0t}( x_t|x_0,  y) = \mathcal{N}\left(x_t; \mu_t(x_0, y), \sigma_t^2 {I}\right).
\end{equation}
We call $\mu_t(x_0, y)$ the \textit{mean-evolution} and $\sigma_t^2$ the \textit{variance-evolution} as they describe how the mean and variance of the process state $x_t$ evolve over the diffusion time $t$. If we can find closed-form solutions for the mean and variance evolution, then \eqref{eq:perturbation-kernel} allows us to efficiently compute the process state $x_t$ for each $t$ by calculating
\begin{equation} \label{eq:eff_sampling1}
    x_t = \mu_t( x_0,   y) + \sigma_t z,
\end{equation}
with $z \sim \mathcal N(0,1)$.

\subsection{Interpolating SDEs: Unifying Conditional Diffusion} \label{sec:isde}
In this work, we are interested in \acp{isde}. These are \acp{sde}, whose mean-evolution interpolates between the degraded observation and the clean speech: 

\begin{align} 
    \mu_t(x_0, y) &= (1-k(t)) x_0 + k(t) y \label{eq:mean_inter} \\
    &=  x_0 + k(t)(y - x_0) \label{eq:mean_inter2}
\end{align}
with $0 \leq k(t) \leq 1$ being a monotonically increasing continuous interpolation function, and $k(0)=0$ and $\lim_{t \to \Tmax} k(t)=1$. Eq. \eqref{eq:mean_inter2} shows that gradually the difference $y - x_0$ is added to the clean speech signal $x_0$. This difference has different meanings for different audio tasks. For example, in Speech Enhancement/Noise reduction, the difference $y-x$ corresponds to environmental noise that has to be removed. This means that for Noise reduction, $\mu_t$ is a noisy mixture depending on the corrupted signal $y$, but with a higher \ac{snr}. For \ac{bwe}, the difference $y-x$ corresponds to the high-frequency content that has to be restored.

As a novelty, here we unify the formulations of different \acp{isde} from popular approaches such as in \cite{lay202interspeech, vpidm, trachu24_interspeech, qiu2023sebridge, jun2025snraligned, lee2025flowse, jukic2024sbse}. 
We show in the Appendix (\cref{sec:appendix:proof1}) that all \acp{isde} must have the following drift coefficient:
\begin{equation} \label{eq:interpol-drift}
    f_t(x_t, y) = \gamma(t) (  y -  x_t ) 
\end{equation}
with $\int_{0}^t \gamma(s)ds \to \infty$ for $t \rightarrow \Tmax$, and  $\gamma(t) \geq 0$. We call $\gamma(t)$ the stiffness function of the \ac{isde}. The converse is also true and can be computed from \cite[(6.10)]{kara_and_shreve}, meaning if \eqref{eq:interpol-drift} is the drift coefficient, then the resulting mean-evolution is interpolating between $x_0$ and $y$.

Moreover, we can even connect the interpolation function $k(t)$ to the stiffness factor $\gamma(t)$. We have that the stiffness parameter $\gamma(t)$ and the interpolation function $k(t)$ are connected via the following equations:
\begin{equation} \label{eq:inter_stiff_1}
    \gamma(t) = \frac{k'(t)}{1-k(t)}, 
\end{equation}
where $k'(t)$ denotes the derivative of $k(t)$ with respect to $t$. Solving for $k(t)$ in \eqref{eq:inter_stiff_1}, we can equivalently write
\begin{equation} \label{eq:stiff_inter_2}
    k(t) = 1-e^{- \int_{0}^t \gamma(s)ds}.
\end{equation}

In the following, we provide a recipe for how to find drift and diffusion coefficients of a linear \ac{sde}, given an interpolating mean with interpolation function $k(t)$ and given the standard deviation $\sigma_t$. Given an interpolating mean, we can already find the drift coefficient from \eqref{eq:inter_stiff_1}. If the diffusion coefficient $g(t)$ in \eqref{eq:fsde} is given, then in the case of linear \acp{sde}, the variance $\sigma_t^2$ can be computed from \cite[Eq. (6.11)]{kara_and_shreve}:
\begin{equation}\label{eq:compute_var}
\sigma_t^2 = e^{-2\int_0^t \gamma(s) \D s}\int_0^te^{2\int_0^u \gamma(s) \D s} g(u)^2 \D u.
\end{equation}
 Conversely, if one wants to design an interpolating \ac{sde} process with a desired variance, we can solve \eqref{eq:compute_var} for $g(t)$ to obtain:
 \begin{equation}\label{eq:compute_diff}
    g^2(t) = \frac{\D (\sigma_t^2 \cdot e^{2\int_0^t \gamma(s) \D s})}{\D t}  e^{-2\int_0^t \gamma(s) \D s}.
 \end{equation}
Therefore, for a given process defined by its mean $\mu_t$ and variance $\sigma_t^2$ evolution, we find a linear \ac{sde} \eqref{eq:fsde} with the desired mean and variance by computing the drift coefficient $f_t$ in \eqref{eq:interpol-drift} from  \eqref{eq:inter_stiff_1}, and the diffusion coefficient $g(t)$ from \eqref{eq:compute_diff}.

In \cref{tab:sde}, we provide a list of popular \acp{isde} for \ac{sr} tasks, defined by their diffusion coefficient $g(t)$ and drift coefficient $f_t$. By \eqref{eq:interpol-drift}, for the drift coefficient $f_t$, it is enough to state the stiffness factor $\gamma(t)$. Hence, in \cref{tab:sde}, we only show $\gamma(t)$.
The resulting closed-form solution of the mean and variance-evolution are indicated by the interpolation function $k(t)$ and the standard deviation $\sigma_t$. It is interesting to see that if $\Tmax < \infty$, the drift coefficient becomes numerically unstable, as it can be seen for BBED and Optimal Transport. In fact, this is a general implication of \acp{isde} which can be explained as follows.

By condition written below \eqref{eq:mean_inter2}, we have $\lim_{t \to \Tmax} k(t)=1$. Hence, from \eqref{eq:stiff_inter_2}, we must have that exponential term becomes 0. This implies, $\int_{0}^t \gamma(s)ds$ approaches $\infty$, when $t \to \Tmax$. Therefore, $\gamma(t)$ diverges as $t\to\Tmax$, which leads to numerical instability near the terminal time. We can evade the numerical instability for $\gamma(\Tmax)$, if we construct an \ac{sde} such that the last diffusion time-step is unbounded, i.e. $\Tmax = \infty$. Such an \ac{sde} is given by the \ac{ouve} \ac{sde} in \cref{tab:sde}. One design issue of the \ac{ouve} is that the introduced parameters $\smax, \smin$ do not follow what their name suggests. That is, $\smax$ (or $\smin$) is not the maximal $\sigma_t$ (or minimal $\sigma_t$). To fix this issue, we construct \ac{fouve} SDE. To this end, we set the drift coefficient as for OUVE, but the desired standard deviation is $\smin(\frac{\smax}{\smin})^{t}$. The corresponding diffusion coefficient can be computed from \eqref{eq:compute_diff} and its result provided in \cref{tab:sde}. For fOUVE we have now that the maximal $\sigma_t$ is indeed given by $\smax$, likewise $\smin$ is $\min\limits_{t} \sigma_t$. Therefore, fOUVE matches the intuitive meaning of $\smin,\smax$, and facilitates a grid-search of $\smax, \smin$ as discussed in \cref{sec:exp_setup:isde}.

Next, we want to derive a fast sampler for the class of introduced \acp{isde}. To this end, we introduce the reverse process and the corresponding \ac{pfode} and reverse \ac{sde} in the next section. In addition, in \cref{sec:rk}, we explain fast solvers from the literature and how to apply them to the proposed class of \acp{isde} in \cref{sec:rk:isde}.

\subsection{Reverse Process}
Under mild constraints \cite{anderson1982reverse, Zhang2022FastSO}, a (forward) \ac{sde} can be reversed in time. This means there exists a reverse \ac{sde} starting in $t=\Tmax$, and ending in $t=0$, that has the same marginals 
as its forward \ac{sde} \cite{song2021sde}. In fact, \cite{Zhang2022FastSO} shows that there even exists a family of reverse \acp{sde} parameterized by $0\leq \kappa \leq 1$:
\begin{align}\label{eq:family_rsde}
    \D{  x_t} =&
        \left[
            f_t(  x_t,  y) - \frac{1+\kappa²}{2}g(t)^2  \nabla_{  x_t} \log p_t(  x_t| y)
        \right] \D{t} \\ \notag
        &+ \kappa g(t)\D{\bar{ w}}\,, 
\end{align}
where
$\D{\bar{ w}}$ is a Wiener process going backwards in time. For $\kappa=1$, we obtain the reverse \ac{sde} by Anderson \cite{anderson1982reverse}, and for $\kappa=0$, we obtain the so-called \ac{pfode}.

To solve this reverse \ac{sde}, we must set a finite $T \lesssim \Tmax$. This is because, in the case of finite $\Tmax$, the drift coefficient $f(x_t,y)=\gamma(t)(y-x_t)$ becomes numerically unstable as $\gamma(t)\to \infty$, if $t\to\Tmax$ as discussed in \cref{sec:isde}. Obviously, in the case $\Tmax = \infty$, we must also set a finite $T<\Tmax$. We start sampling from an initial $x_T = \mu_T(x_0,y) + \sigma_{\Tmax}z$, since $\mu_T(x_0, y)$ depends on $x_0$ which is unknown during inference, we simply approximate $\mu_T(x_0, y) \approx y$ assuming that $T$ is sufficiently close to $\Tmax$. The score $\score$ is unknown during inference, and is typically approximated by a \ac{nn}. To approximate the score, we can train on the \ac{dsm} loss. That is, given $x_t = \mu_t(x_0,y) + \sigma_tz$, with $z \sim \mathcal N(0,1)$, and $\nabla_{  x_t} \log p_t(  x_t| y, x_0) = -\frac{\epsilon}{\sigma_t}$, we optimize on:

\begin{equation}\label{eq:dsm}
     \Ldsm = \mathbb{E}_{t,( x_0, y),  \epsilon,  x_t|(x_0,y)} \left[
        \norm{ \scorem(x_t,  y, t) + \frac{ \epsilon}{\sigma_t}}_2^2
    \right]\,.
\end{equation}
Another option is to directly estimate $\epsilon$ with a \ac{nn} $\epsm$ which yields the $\epsilon$ loss:
\begin{equation}\label{eq:eps}
     L_{\epsilon} = \mathbb{E}_{t,( x_0,  y), \epsilon, x_t|( x_0,y)} \left[
        \norm{ \rho_\theta( x_t,  y, t) - \epsilon}_2^2
    \right]\,.
\end{equation}
With a trained $\epsm$, we can approximate $\score \approx -\frac{\epsm(x_t,  y, t)}{\sigma_t}$. In the following, to simplify notation, we use $\hat{s}_\theta(x_{t}, y, t) \approx \nabla_{x_t} \log p_t(x_t| y)$  to comprise both $\scorem$ and $-\frac{\epsm(x_t,  y, t)}{\sigma_t}$.

\section{Runge-Kutta ODE Solvers} \label{sec:rk}
In this section, we first set $\kappa = 0$ in \eqref{eq:family_rsde} to particularly address \ac{ode} solvers as they are well-established in the literature \cite{butcher2016numerical}. In \cref{sec:rk:isde}, we present our proposed to solution to address a wide class of iSDEs with $0 \leq \kappa \leq 1$.

Equipped with a trained \ac{nn} estimating the score, we fix a reverse diffusion-time schedule:
$T = t_N > \dots > t_1 > t_0 = 0$. Then, starting with $x_T$, for each diffusion time-step from $\ti$ to $\timo$, we aim to solve the integral:

\begin{equation} \label{eq:integralform}
    x_{\timo} = x_{\ti} +  \int_{\timo}^{\ti}  f_{t}(  x_{t},  y) - \frac{1}{2}g(t)^2 \hat{s}_\theta(x_{t}, y, t) \D t
\end{equation}
A widely used method is the \ac{rk} method. For \ac{rk}, the integral \eqref{eq:integralform} is solved by evaluating the integrand at $p \in \mathbb{N}$ specific intermediate points, yielding a local truncation error of $\mathcal{O}((\ti -\timo)^{p+1})$. For instance, the Midpoint method evaluates the integral at each time-step ($\ti$ to $\timo$) at $\ti$ and additionally at the midpoint $\frac{\ti+\timo}{2}$. Therefore, this method has 2 evaluations of the integrand per time-step. The Midpoint method is an RK2 method and has a cubic local truncation error.

Instead of directly evaluating the integral at intermediate points, we can integrate the linear term $f_{t}(x_{t},  y)$ out of the integral to obtain
\begin{align} \label{eq:linear}
    L(x_{\ti}, y, \ti, \timo) & \coloneqq  x_{\ti} + \int_{\ti}^{\timo} f_t(x_t,y) \D t \nonumber \\
    &=\Psi(\timo, \ti) x_{\ti} + \left( 1-\Psi(\timo, \ti)\right)y, 
\end{align}
where for $s > t$,
\begin{equation}
    \Psi(s,t) = e^{-\int_{s}^{t}\gamma(\tau)\D \tau} = \frac{1-k(t)}{1-k(s)}
\end{equation}
is the so-called fundamental solution of the homogeneous equation. As the linear term is exponentially weighted by $e^{-\int_{s}^{t}\gamma(\tau)\D \tau}$, we refer to this solver as \ac{erk}. We define the integral over the second term containing $\hat{s}_\theta$ in \eqref{eq:integralform} as the non-linear part. The non-linear part can be rewritten as
\begin{equation} \label{eq:nonlinear}
    N(\ti, \timo) = \int_{\timo}^{\ti} \Psi(\tau, \ti) \frac{1}{2}g^2(\tau) \hat{s}_\theta(x_{\tau}, y, \tau) \D \tau.
\end{equation}
In summary, we can write the solution
\begin{equation} \label{eq:split_linear}
    x_{\timo} =  L(x_{\ti}, y, \ti, \timo) + N(\ti, \timo),
\end{equation} 
which separates the solution into a linear and a non-linear part. Since the linear part $L(x_{\ti}, y, \ti, \timo)$ is exactly integrated in \eqref{eq:linear}, we are left with integrating the non-linear part \eqref{eq:nonlinear}. 
The DPM-Solver \cite{Lu2022DPMSolver}, abbreviated DPM-pS, is a solver that follows the scheme of separating the linear from the non-linear term, and has been proposed for unconditional diffusion processes (meaning $y=0$) and has local truncation error of order $p$. Briefly explained, DPM-pS proposes two ideas for how to integrate the non-linear part: 1) Taylor expand $\hat{s}_\theta$, and pull it out of the integral, and 2) simplify the integral of the remaining non-linear part with a substitution rule, which we will explain in more detail in the following.

\subsection{Existing Fast DPM-Solver}
In this section, we explain the fast \ac{ode} sampler, called DPM-pS \cite{Lu2022DPMSolver}, which is based on \eqref{eq:split_linear} with $y=0$. Therefore, it can be understood as a direct application of \ac{erk} \cite{lai2025principles}. The linear term then results in  
\begin{equation} \label{eq:dpm:linear}
    L(x_{\ti}, \ti, \timo) = \frac{1-k(\timo)}{1-k(\ti)} x_{\ti}.
\end{equation}
In the notation from \eqref{eq:dpm:linear}, we remove $y$ from the input of $L$, whenever it is independent of $y$ (i.e. $y=0$). Computing the non-linear part is more problematic than computing the linear part, as we need to integrate over the \ac{nn}. For DPM-pS, the $\epsilon$-loss is employed, together with the change-of-variables $\lambda_t = \log(\frac{1-k(t)}{\sigma_t})$ the non-linear part from \eqref{eq:nonlinear} becomes 

\begin{equation} \label{eq:dpm_solver:non_linear_simple}
    N(\ti, \timo)=(1-k(\ltimo))\int_{\lti}^{\ltimo} e^{-\lambda} \rho_\theta(x_{\lambda}, \lambda)
\end{equation}
To avoid integration of the \ac{nn}, the $(p-1)$-th degree Taylor power series is employed as:
\begin{equation}
     \rho_\theta( x_\lambda, \lambda) \approx \sum_{n=0}^{p-1} \frac{(\lambda-\lti)^{n}}{n!}\epsm^{(n)}(x_{\lti}, \lti), 
\end{equation}
where $\epsm^{(n)}$ refers to the $n$-th total derivative of the model $\epsm(\cdot)$. Replacing the \ac{nn} in the integral of the non-linear part with the Taylor power series yields:
\begin{align}\label{eq:dpm_solver:nonlinear}
    N_p(x_{\ti}, \ti, \timo) \coloneqq &(1-k(\ltimo)) \cdot R_p(x_{\ti}, \ti, \timo)\\
    R_p(x_{\ti}, \ti, \timo) \coloneqq  & \sum_{n=0}^{p-1} 
    \epsm^{(n)}(x_{\lti}, \lti) w_n(\lti, \ltimo),
\end{align}
with $N_p(x_{\lti}, \lti, \ltimo) = N(\lti, \ltimo) + \mathcal{O}((\ti - \timo)^{p})$. In addition, the weights $w(\lti, \ltimo)$ are given by
\begin{equation}\label{eq:dpm_solver:weights}
    w_n(\lti, \ltimo) = \int_{\lti}^{\ltimo} e^{-\lambda} \frac{(\lambda-\ltimo)^n}{n!} \D \lambda.
\end{equation}
A key observation of \cite{Lu2022DPMSolver} is that the change-of-variable trick introduced by $\lambda_t$ eliminates direct dependencies on the diffusion and drift coefficients, as these coefficients do not directly appear anymore in the integral form in \eqref{eq:dpm_solver:weights}. Consequently, the integral in \eqref{eq:dpm_solver:weights} can be solved in closed form. Therefore, solving \eqref{eq:dpm_solver:nonlinear} only boils down to computing the $n$-th total derivative $\epsm^{(n)}(x_{\lti}, \lti)$ for which the authors of \cite{Lu2022DPMSolver} refer to literature \cite{hochbruckExpRK2005, LuanEffExpRK2021}. The whole (truncated) form can be written as 
\begin{equation} \label{eq:dpm_solver}
    x_{\timo} \approx  L(x_{\ti}, \ti, \timo) + N_p(x_{\ti}, \ti, \timo).
\end{equation}

\subsection{Proposed Fast ISDE-Solver} \label{sec:rk:isde}
As a contribution, we derive a fast solver similar to DPM-pS for the proposed wide class of \acp{isde} (with some examples in \cref{tab:sde}). This novel solver solves \eqref{eq:family_rsde} with $0 \leq \kappa \leq 1$. There are two important distinctions to make. First, the drift coefficient depends on the noisy observation $y \neq 0$. Second, often the \ac{dsm} loss is employed \cite{lay202interspeech, richter_sgmse} instead of the $\epsilon$ loss for \ac{sr} tasks. Since the drift coefficient depends on $y$, the linear term is now of the form:
\begin{align} \label{eq:isde_solver:linear}
   L(x_{\ti}, y, \ti, \timo) =&  \frac{1-k(\timo)}{1-k(\ti)}x_{\ti}  \notag \\ &+ \left(1-   \frac{1-k(\timo)}{1-k(\ti)}\right)y.
\end{align}
Compared to \eqref{eq:dpm:linear}, we see that this linear term also includes a term depending on $y$. For the non-linear part and in contrast to DPM-pS, the change-of-variables $\lambda_t = \log\left(\frac{1-k(t)}{\sigma_t}\right)$ does not eliminate the dependence on $\sigma_t$ when the score is directly approximated using $\score \approx \scorem$ opposed to $\score \approx -\frac{\epsm}{\sigma_t}$.
However, following the idea of using a Taylor power series, we can pull the \ac{nn} out of the integral:

\begin{align}\label{eq:isde_solver:nonlinear}
    N_p^\kappa(x_{\ti}, \ti, \timo) \coloneqq & (1-k(\timo)) \cdot R_p^\kappa(x_{\ti}, \ti, \timo)\\
    R_p^\kappa(x_{\ti}, \ti, \timo) \coloneqq  &  (1 + \kappa^2)\sum_{n=0}^{p-1} 
    \scorem^{(n)}(x_{\ti}, y, \ti) w_n(\ti, \timo).
\end{align}
As before, this truncated non-linear form approximates the true non-linear part as $N_p^\kappa(x_{\ti}, \ti, \timo) = N(\ti, \timo) + \mathcal{O}((\ti - \timo)^{p})$. However, the weights differ now from \eqref{eq:dpm_solver:weights}:
\begin{equation} \label{eq:isde_solver:weights}
      w_n(\ti, \timo) = \int_{\ti}^{\timo} \frac{g(\tau)^2}{2(1-k(\tau))}  \frac{(\tau - \ti)^{n}}{n!} d\tau.
\end{equation} 
While for DPM-pS, $w_n$ is solved in closed form for any choice of diffusion and drift coefficient, for the proposed solver, this is not generally possible. This is because the integrand includes the standard deviation $\sigma_t$, which depends on the diffusion and drift coefficients. In those cases, one can make use of numerical integral solvers \cite{quadpack} whose computational expenses are very small compared to computing the $n$-th total derivatives of the \ac{nn} $\scorem^{(n)}$. However, for certain choices of the drift and diffusion coefficients, $w_n$ can be solved analytically. In fact, in Appendix~\ref{sec:weights_fouve_ouve}, we derived closed-form solutions for fOUVE and OUVE. Unlike DPM-pS, we allow $0 \leq \kappa \leq 1$. Therefore, the solution $x_{\timo}$ after taking one step from $\ti$ to $\timo$ can be written as
\begin{equation} \label{eq:isde_solver}
    x_{\timo} \approx  L(x_{\ti}, y, \ti, \timo) + N_p^\kappa(x_{\ti}, \ti, \timo) + \kappa I(\ti, \timo)z,
\end{equation}
where $zI(\ti, \timo)$ is the exactly integrated It\^o-integral injecting random Gaussian noise $z\sim \mathcal N(0,1)$. The computations of this It\^o-integral for fOUVE can be found in the Appendix \ref{sec:appendix:itointegral_fouve_ouve}. We call the resulting solver iSDE-pS-$\kappa$, where, as before, $p$ refers here to the truncation error, and $\kappa$ regulates how much Gaussian noise is injected at each time-step. For $\kappa=0$, this solver solves the \ac{pfode} and follows as the DPM-pS the \ac{erk} formulation in \eqref{eq:split_linear}. For $0 < \kappa \leq 1$, this solver solves a reverse \ac{sde}.

In \cref{alg:fast_isde_solver_2}, we show such a solver for $p=2$. In line 4 of \cref{alg:fast_isde_solver_2}, the solution at the midpoint $s_i$ is computed with $p=1$ which requires the evaluation of $\scorem(x_{\ti},y,\ti)$ in $N_1^0(x_{\ti}, y, \ti)$. Another score-model evaluation is due to line 5, as we compute $\scorem(x_{s_i},y, s_i)$. Both score-model evaluations are then reused in line 6 for computing $N_2(x_{\ti}, y, \ti)$. Therefore, the \ac{nfe} is $2N$, as we have $N$ time-steps, and 2 \ac{nfe} per time-step. In the last line, we add Gaussian noise $z \sim \mathcal N(0,1)$ to the process if $\kappa > 0$.

\begin{algorithm}
\caption{iSDE-2S-$\kappa$} \label{alg:fast_isde_solver_2}
\begin{algorithmic}[1]
\Require $x_T$, time-steps $T = t_N> \dots t_1=0$, score model $\scorem$
\State $x_{t_i} \leftarrow x_T$
\For{$i = N,  \dots, 0$}
    \State $s_i \leftarrow \frac{\ti + \timo}{2}$
    \State $x_{s_i} \leftarrow L(x_{\ti}, y, \ti, s_i) + N_1^0(x_{\ti}, \ti, s_i)$ 
    \State $\tilde{\scorem}^{(1)} \leftarrow \frac{\scorem(x_{\ti}, y, \ti)-\scorem(x_{s_i}, y, s_i)}{2(\ti - \timo)}$ \Comment{Approx. $\scorem^{(1)}$}
    \State $x_{\timo} \leftarrow L(x_{\ti}, y, \ti, \timo) + N_2^\kappa(x_{\ti}, \ti, \timo) $
    \State $x_{\timo} \leftarrow x_{\timo} + \kappa I(\ti,\timo)z$  \Comment{$z \sim \mathcal N(0,1)$}

\EndFor
\end{algorithmic}
\end{algorithm}

\section{Experimental Setup} \label{sec:exp_setup}
\subsection{Data representation}  \label{sec:exp_setup:data}
Each audio input, sampled at 16 kHz, is converted to a complex-valued \ac{stft} representation. As in \cite{diffusionbuffer}, we use a window size of $\nfft =510$ samples ($\approx \qty{32}{\ms}$), a hop length of $\nhop=256$ samples ($\qty{16}{\ms}$), and a periodic Hann window. The input for training is cropped to $ K=128$ time frames randomly, resulting in approximately 1 second of data. A magnitude compression is used to compensate for the typically heavy-tailed distribution of \ac{stft} speech magnitudes~\cite{gerkmann2010empirical}. Each complex coefficient $v$ of the \ac{stft} representation is transformed as $\beta |v|^\alpha \mathrm e^{i \angle(v)}$. Depending on the task, we use different $\alpha, \beta$ which are reported in \cref{sec:exp_setup:isde}.

\subsection{Data set and audio tasks} \label{sec:exp_setup:dataset}
\ac{sr} is the task of retrieving the clean speech signal from a degraded signal. Although the diffusion process is formulated in the \ac{stft} domain, we define the following corruption operations in this section in the time-domain. To this end, we denote with $\ts, \ty$ time-domain signals of the clean speech signal and corrupted signal. In  general, we can write 
\begin{equation}
    \ty = u(\ts),
\end{equation}
where $u(\cdot)$ is the corruption operator. The precise definition of $u$ depends on which corruption type or audio task is considered. In this work, we investigate the following audio tasks: Noise reduction, Dereverberation, Declipping, MP3 decoding, and \ac{bwe} which we describe next.

\subsubsection{Noise reduction} \label{sec:exp_setup:dataset:se}
In Noise reduction, the corruption operator adds environmental noise to the clean speech signal. We train, validate, and test on the publicly available dataset EARS-WHAM-v2\footnote{\url{https://github.com/sp-uhh/ears\_benchmark}} with clean files from the EARS dataset \cite{richter2024ears} and environmental noise files from the WHAM dataset \cite{wham}. The dataset, originally recorded at 48 kHz, was downsampled for this work to 16 kHz. This dataset has 54 hours for training, 1.1 hours of validation, and 3.5 hours for testing. 

\subsubsection{Bandwidth Extension}
For \ac{bwe} we use the same split and clean files at 16\,kHz from EARS-WHAM-v2 where the corrupted files are obtained by downsampling to sampling frequencies of 8\,kHz and 4\,kHz, leading to a frequency cutoff at 4\,kHz and 2\,kHz, respectively. This yields, as before, 54 hours for training, 1.1 hours of validation, and 3.5 hours of testing.

\subsubsection{Dereverberation}
We investigate speech dereverberation using the EARS-Reverb-v2 dataset \cite{richter2024ears}. The signal corruption model is $\ty = \ts * h$, where $*$ indicates time-domain convolution and $h$ is a sampled \ac{rir}.

\subsubsection{MP3 Decoding}
We investigate the restoration of speech signals degraded by lossy MP3 compression. The corruption operator $u(\cdot)$ degrades the clean speech signal into the MP3 format and subsequently introduces quantization artifacts, band-limitation and other artifacts. We artificially degrade the clean utterances from the EARS-WHAM-v2 dataset using the audiomentations library in Python with a uniformly random bitrate of $\{16, 24, 32, 40, 48, 56, 64 \}$ kbp/s. The dataset splits remain identical to the previous tasks, yielding 54 hours for training, 1.1 hours for validation, and 3.5 hours for testing at 16 kHz.

\subsubsection{Declipping}
The corruption operator $u(\cdot)$ applies hard clipping to the clean waveform as
\begin{equation}
\ty[n] =
\begin{cases}
\ts[n], & \text{if } -\xi< \ts[n] < \xi \\
\xi \cdot \text{sign}(\ts[n]), & \text{otherwise}
\end{cases}
\end{equation}
where $y[n]$ is the $n$-th sample of $\ty$. In addition, $\xi$ is the clipping threshold. The threshold and clean files are chosen as follows. As before, we use the same split and clean files from EARS-WHAM-v2. We multiply the peak normalized clean speech file by a uniformly random number $g \in [0.3, 1.0]$. The threshold is then uniformly randomly selected $\xi \in [0.05g, 0.3g]$. With these parameters, $0.11$ percent of the samples are clipped.

\subsection{Metrics}\label{sec:exp:metrics}
To evaluate the performance of the proposed method, we use standard metrics, which we will describe in detail below. \\
\textbf{PESQ:} The Perceptual Evaluation of Speech Quality (PESQ) is used for objective speech quality testing and is standardized in ITU-T P.862 \cite{rixPerceptualEvaluationSpeech2001}. The PESQ score takes values between 1 (poor) and 4.5 (excellent). We use the wideband PESQ version. \\
\textbf{DistillMOS:} DistillMOS \cite{stahl2025distillation} is a \ac{mos} prediction method, built by distilling a wav2vec2.0-based \ac{SQA} model into a more efficient model. Higher values indicate better quality.\\
\textbf{SI-SDR:} 
Scale-Invariant-Signal-to-Distortion Ratio (SI-SDR) is a standard evaluation metric for single-channel speech enhancement and speech separation \cite{leroux2018sdr}, which is measured in dB, therefore higher values are better. \\
\textbf{LSD:} The log spectral distance (LSD) metric is like the SI-SDR, a sample-based metric. In contrast to SI-SDR this metric is a distance metric, this means we have that lower values correspond to a better performance.\\
\textbf{FADTK:} Fr{\'e}chet-Audio-Distance metric \cite{Kilgour2019FrchetAD} measures the similarities of distribution of clean speech to the distribution of restored files. Lower values indicate that the distribution of restored files is closer to the distribution of the clean speech signals.

\subsection{Training and Parameter of ISDE}\label{sec:exp_setup:isde}
We train fOUVE on the different tasks. 
For each task, we first select the parameter $\alpha, \beta$ of the magnitude compression. Then, we select the parameters of fOUVE $\smax, \smin, \gamma_0$ report the numbers below. To find an initial $\smax$ and $\beta$, we follow \cite{welker2025flowdec}. We first fix a $\alpha = 0.5$. Then we select $\beta$ as follows.
We ensure that 99.7 percent of the imaginary and real parts of the clean data $s$ remain within $[-1, 1]$. Denote with $s_m, y_m$ the magnitude compressed STFTs of clean and noisy, respectively. We then select an initial $\smax$ , called $\smax_{,0}$, by computing the RMSE between $s_m$ and $y_m$, and take the $q$-th quantile ($q=0.997)$ of all RMSEs of all data points from the data sets. We set $\smin = 0.001$ and $\gamma_0 = 2$. Based on these initial choices of parameters, we further grid-search on $\{\smax_{,0}, 2\smax_{,0}, 3\smax_{,0}\}$ for an optimal maximum $\sigma_t$ for each task. With this scheme, we find for Noise reduction, we have $\beta = 0.26$, and $\smax_{,0} = 0.10$. For \ac{bwe}, we have $\beta = 0.23$, and $\smax_{,0} = 0.07$. For Dereverberation, we have $\beta = 0.23$, and $\smax_{,0} = 0.12$. For MP3 decoding, we have $\beta = 0.24$, and $\smax_{,0} = 0.05$. For Declipping, we have $\beta = 0.1$, and $\smax_{,0} = 0.08$. 

From the three trained models (with three different $\smax$), we select only one model for testing. 
To select a model for testing for each task, we select 10 pairs of clean and degraded signals from the validation set during each training run. These 10 files are then enhanced by solving the reverse \ac{sde} in \eqref{eq:family_rsde} with $\kappa=1$ with \ac{eum} with $M=60$ steps. Based on these enhanced files from the validation set, we selected the best run in terms of SI-SDR for testing.

We use the 2D-UNet NCSN++ as a backbone trained on the $\Ldsm$ loss as it was done in \cite{richter_sgmse, lay202interspeech} and optimized with ADAM \cite{kingma2015adam} with a batch-size of $16$. To avoid numerical inaccuracy of the $\Ldsm$ loss around $0$, we uniformly sample the diffusion time-step 
$t \in [\delta, T]$ with $\delta=10^{-2}$ and $T=1$ for fOUVE as indicated in \cref{tab:sde}.

    \label{fig:specs_bwe}

    \label{fig:specs_mp3}

\subsection{Methods} \label{sec:exp_setup:methods}
We compare against different \ac{ode} and \ac{sde} solvers. To this end, we fixed reverse time-steps $T=t_M > \dots > t_2 = \delta > t_1=0$. \\
\textbf{iSDE-2S-$\kappa$:} We employ the proposed sampler from \cref{alg:fast_isde_solver_2}. This sampler has 2 \acp{nfe} per time-step, resulting in $2M$ \acp{nfe}. \\
\textbf{Euler-Maruyama:} A widely used first-order \ac{sde} solver. Often this solver solves \eqref{eq:family_rsde} with $\kappa=1$ \cite{richter_sgmse, lay202interspeech}. We denote this solver by \ac{eum}. This solver has $M$ \acp{nfe}. \\
\textbf{PC-Sampler:} As for \ac{eum}, this scheme solves \eqref{eq:family_rsde} with $\kappa=1$ as it has been done in \cite{richter_sgmse}. For a time-step from $\ti$ to $\timo$ the \ac{pc} scheme \cite{song2019generative} predicts $x_{\timo}$ by \ac{eum}, and additionally one correction step by annealed Langevin Dynamics resulting in $2M$ \acp{nfe}. In \cite{richter_sgmse} it has been found that it is beneficial to employ one corrector step with a corrector stepsize of $0.5$.\\
\textbf{RK2 (midpoint):} We also employ the Midpoint method, which is a \ac{rk}2 method. This method also has $2M$ \acp{nfe} and is very comparable to the proposed solver iSDE-2S-$\kappa$ as both solvers are second-order, when $\kappa=0$. \\
\textbf{adaptive RK45:} We also employ the "classic RK45" method which is a \ac{rk}4 method. We abbreviate this method with adaptive RK45. An adaptive method does not employ a fixed time-step schedule, instead, the sampler automatically adapts the number of steps of a time-step schedule to obtain a new time-step schedule with more steps. Then the sampler compares the two different time-step schedules and their difference. If the difference does not decrease below a certain threshold (given by parameters $\rtol=10^{-5}, \atol=10^{-5}$), then the sampler will stop the comparisons of different schedulers. Adaptive methods aim to take the least number of steps while being numerically optimal.

\begin{figure*}[t]
\vspace{-0.0cm}
    \centering
    \includegraphics[width=1.0\linewidth]{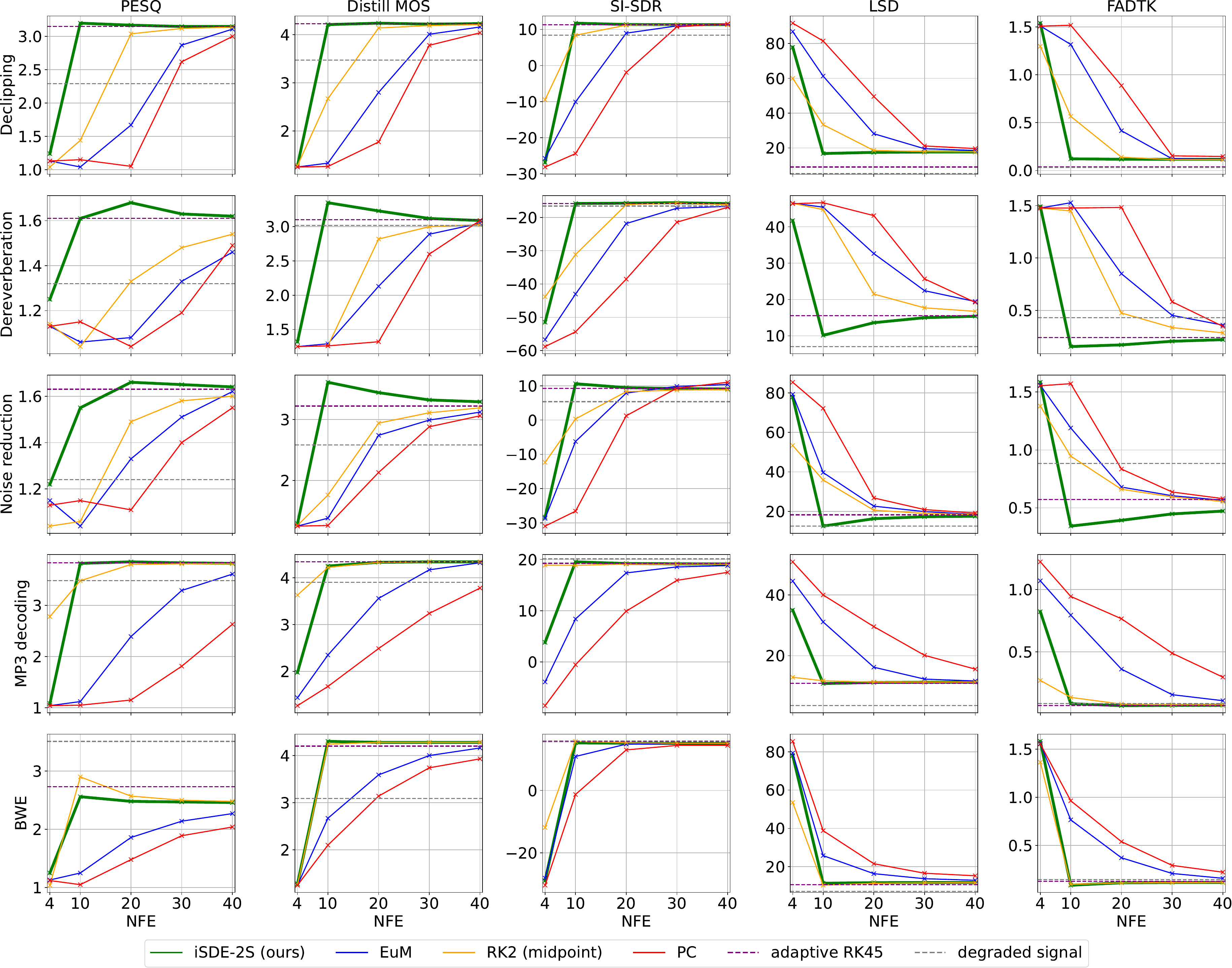} 
    \caption{Results on the different tasks with different samplers. For all tasks, adaptive RK45 uses more than 40 \acp{nfe}.}\label{tab:audio_tasks}
\end{figure*}

\section{Results} \label{sec:res}
In \cref{sec:res:audio_tasks}, we discuss the results of the proposed iSDE-2S-$\kappa$ solver with $\kappa = 0$, for which we simply write iSDE-2S for brevity. In \cref{sec:res:kappa} we see the effect of varying $\kappa$. 

\subsection{iSDE-2S vs other solvers} \label{sec:res:audio_tasks}
For a fair comparison of the solvers, we ensure that the solvers use the same \acp{nfe} from which we set the number of equally spaced time-steps $M$.

In \cref{tab:audio_tasks}, we report the results of the proposed iSDE-2S solver compared against other solvers as described in \cref{sec:exp_setup:methods} on several audio tasks described in \cref{sec:exp_setup:dataset}. We observe that iSDE-2S outperforms all other solvers at \ac{nfe}=10 for Declipping, Dereverberation, and Noise reduction. If the \ac{nfe} is larger than 10, then the other solvers may reach the same performance as the proposed iSDE-2S. For instance, for Dereverberation, the solvers EuM, RK2 (midpoint), and PC require 40 \ac{nfe} to achieve virtually the same performance as the proposed iSDE-2S solver with only 10 \ac{nfe} in DistillMOS and SI-SDR. Moreover, in PESQ remains a relatively large gap of at least $0.08$ between the proposed solver and EuM, RK2 (midpoint), and PC. In addition, adaptive RK45 even uses 91 \acp{nfe} for Dereverberation. This demonstrates that the proposed iSDE-2S restores the clean speech signal among different audio tasks, much faster than usual solvers, with better metrics.

There are two exceptions in which the proposed iSDE-2S does not outperform the RK2 (midpoint) method with fewer \acp{nfe}. First, on \ac{bwe}, the Midpoint method performs on par with the proposed iSDE-2S solver for all metrics and \acp{nfe}. Notably, there is a difference in PESQ at \ac{nfe} = 10. However, it should be mentioned that PESQ does not reflect well if a low-pass filtered signal has been correctly bandwidth-extended \cite{pesqisbad}.
Secondly, for MP3 corruption, we also observe that the RK2 (midpoint) method is on par with the proposed iSDE-2S except for \acp{nfe} = 4, where Midpoint outperforms the proposed iSDE-2S in terms of metrics. The reason why Midpoint with 4 \acp{nfe} has better metrics is that it removes almost all Gaussian noise. However, we report that no MP3 decompression is performed. In all other tasks and for all other solvers with 4 $\acp{nfe}$, we observe that performance is poor as too much residual Gaussian noise is left over. The similar performance of RK2 (midpoint) and iSDE-2S can be explained by the fact that both are second-order solvers for the \ac{pfode}. The key difference is that iSDE-2S integrates the linear term exactly, whereas RK2 approximates it. We believe that both solvers have similar performance if the linear term that has been exactly integrated for iSDE-2S is much less important than the non-linear term. This seems to be the case for fOUVE on \ac{bwe} and MP3 decoding.

As a higher-order solver generally gives a much more accurate solution than a lower-order solver, we have that adaptive RK45 is the strongest solver, as this is a fourth-order solver, and the competing solvers are at most second-order. In addition, this solver has more than 40 \acp{nfe} in all tasks. More precisely, the averaged \acp{nfe} over the test sets for adaptive RK45 is 66, 91, 44, 48, and 75 for Declipping, Dereverberation, Noise reduction, MP3 decoding, and \ac{bwe}, respectively. For all tasks, we see that the maximal performance of the other solvers becomes on par with that of adaptive RK45. However, as discussed before, with the proposed iSDE-2S, this happens with already 10 \acp{nfe}, whereas other solvers may require 40 \acp{nfe}.

\begin{table}
\centering
\begin{tabular}{|l|c|c|c|}
\hline
\textbf{$\kappa$} & PESQ & DistillMOS &  FADTK  \\
\hline
$0$ & $1.55 \pm 0.45$ & $3.61 \pm 0.87$ & $0.34$ \\
$0.05$ & $1.58 \pm 0.47$ & $3.61 \pm 0.88$ & $\mathbf{0.33}$ \\
$0.1$ & $1.67 \pm 0.51$ & $\mathbf{3.63 \pm 0.84}$ & $\mathbf{0.33}$ \\
$0.125$ & $\mathbf{1.73 \pm 0.55}$ & $3.49 \pm 0.77$ & $0.41$ \\
$0.15$ & $1.50 \pm 0.37$ & $3.00 \pm 0.54$ & $0.64$ \\
\hline
\end{tabular}
\caption{Performance of iSDE-2S-$\kappa$ with different $\kappa$ on Noise reduction with $10$ \acp{nfe}.}
\label{tab:kappa}
\end{table}

\subsection{Effect of $\kappa$ for iSDE-2s-$\kappa$} \label{sec:res:kappa}
In \cref{tab:kappa}, we observe the effect of varying $\kappa$ for Noise reduction with 10 \acp{nfe} on fOUVE. More precisely, we use the proposed solver iSDE-2S-$\kappa$, which solves the reverse \ac{sde} \eqref{eq:family_rsde} with the given $\kappa>0$. We can see that increasing $\kappa$ from $0$ to $0.1$ benefits the metrics, whereas in PESQ we observe a difference of over $0.1$ between using $\kappa=0$, and $\kappa = 0.1, 0.125$. A larger $\kappa>0.125$ results in a restored file, where too much Gaussian noise is left. We report that more diffusion time-steps $M$ (or equivalently more \acp{nfe}) are required to remove the injected Gaussian noise when $\kappa$ is too large. In summary, after the score-model has been trained, $\kappa$ can be varied to empirically tune the performance of the proposed solver iSDE-2S-$\kappa$ without additional training.

\section{Conclusion} \label{sec:conclusion}
In this work, we unified the mathematical formulation of popular choices of \acp{sde} for audio tasks. We focused on linear \acp{sde} where the mean-evolution is a linear interpolation between the clean speech and a degraded version of it. Based on these interpolating \acp{sde}, we developed a fast solver, called iSDE-2S-$\kappa$, that solves the probability flow \ac{ode} when $\kappa=0$, and solves reverse \acp{sde} for $0 < \kappa \leq 1$. This work has been inspired by DPM-Solver \cite{Lu2022DPMSolver}, where the underlying \ac{sde} formulation can be understood as a special case for which the degraded signal is $0$, or differently said, DPM-Solver solves for the unconditional probability flow \ac{ode}, and the proposed iSDE-2S-$\kappa$ solves a conditional process. We demonstrated the superiority of the proposed solver against other solvers such as Euler-Maruyama, Predictor-Corrector scheme, RK2 (midpoint) and adaptive RK45. We experimented on Speech Restoration tasks: Declipping, Dereverberation, Noise reduction, MP3 decoding and \ac{bwe}. For MP3 decoding and \ac{bwe}, we find that the proposed solver is on par with RK2 (midpoint), and outperforms Euler-Maruyama and Predictor-Corrector. For Declipping, Dereverberation, and Noise reduction, the proposed solver is much faster, requiring only 10 \acp{nfe} to achieve the same performance as the adaptive RK45 solver that uses much more than 40 \acp{nfe}. At the same time, Euler-Maruyama, Predictor-Corrector, and even RK2 (midpoint) require up to 40 \acp{nfe} to achieve the performance of adaptive RK45, demonstrating the effectiveness of the proposed solver.

\section{Appendix} \label{sec:appendix}
\subsection{Proof \eqref{eq:interpol-drift} and \eqref{eq:inter_stiff_1}} \label{sec:appendix:proof1}
We assume that a linear \ac{isde} is given with an interpolation function $k(t)$. We want to show that \eqref{eq:interpol-drift} and \eqref{eq:inter_stiff_1} hold. 
Every linear \ac{sde} follows the dynamics \cite[(6.12)]{kara_and_shreve}:
\begin{equation}
    \mu_t' = A(t)\mu_t + a(t),
\end{equation}
where $\mu_t'$ is the derivative w.r.t to $t$ of the mean-evolution. Inserting the definition of \acp{isde} from \eqref{eq:mean_inter2} yields:
\begin{equation}
    -k(t)x_0 + k(t)y =A(t) x_0 + (1-k(t))(y-x_0) + a(t)
\end{equation}
Now, collecting terms with $x_0$ on the right-hand side of the equation and comparing them to $-k(t)x_0$ on the left-hand side yields:
\begin{equation}
    A(t) = \frac{-k'(t)}{1-k(t)}
\end{equation}
Likewise, a comparison of the remaining terms of the right-hand side to $k(t)y$ from the left-hand side yields:
\begin{equation}
    a(t) = \frac{k'(t)}{1-k(t)}y.
\end{equation}
Since $f_t(x_t, y) = A(t)x_t+ a(t)$, we can write
\begin{equation}
    f_t(x_t, y) = \frac{k'(t)}{1-k(t)}(y-x_t).
\end{equation}
and from which we conclude that $\gamma(t) =\frac{k'(t)}{1-k(t)}$. We therefore have proven that an \ac{isde} has a drift term of the form \eqref{eq:interpol-drift} and the stiffness function is given by \eqref{eq:inter_stiff_1}.

\subsection{Computing $\omega_n$ for fOUVE and OUVE} \label{sec:weights_fouve_ouve}
In this section, we report the weights $\omega_n$ from \eqref{eq:isde_solver:weights} for the fOUVE and OUVE SDE from \eqref{tab:sde}. For fOUVE and OUVE, we obtain that \eqref{eq:isde_solver:weights} can be written as 

\begin{equation}
    \omega_n(\ti, \timo) = B\int_{\ti}^{\timo} e^{\zeta \tau}  \frac{(\tau - \ti)^{n}}{n!} d\tau
\end{equation}
with $k = \frac{\smin}{\smax}$ and $\zeta = \gamma_0+2\log(k)$. The constant $B = \smin² (\log(k) + \gamma_0)(1+\kappa^2)$ for fOUVE and $B = \smin² \log(k)(1+\kappa^2)$ for OUVE. We can integrate this by parts to obtain

\begin{equation}
     \omega_n(\ti, \timo) = B\left( e^{\zeta \tau}  \frac{(\tau - \ti)^{n+1}}{(n+1)!} \bigg\vert_{\ti}^{\timo} + \frac{1}{\zeta}\omega_{n-1}(\ti, \timo) \right).
\end{equation}
Solving this recursive formulation yields:

\begin{align}
 \omega_n(\ti, \timo) =&  
    B(e^{\zeta \ti }- e^{\zeta \timo })\notag \\ &+B\sum_{i=1}^n\frac{(-1)^{n+i+1}}{\zeta^{n-i+1}}\left(\frac{e^{\zeta \timo}(\timo - \ti)^i}{i!} \right).
\end{align}

\subsection{Computing $I(\ti,\timo)$ for fOUVE} \label{sec:appendix:itointegral_fouve_ouve}
Let $z \sim \mathcal N(0,1)$, then the It\^o-integral can be generally written as a Lebesgue integral as follows:
\begin{align}
        I(\ti, \timo) &=\sqrt{\int_{\timo}^{\ti}  \left(\Phi(\ti, \timo)g(\tau) \right)²\D \tau} \\
         &=(1-k(\timo))\sqrt{\int_{\timo}^{\ti} \left(\frac{g(\tau)}{1-k(\tau)}\right)^2\D \tau},
\end{align}
Inserting $g(t), k(t)$ for fOUVE from \cref{tab:sde}, we obtain
\begin{equation}
    I(\ti, \timo) = (1-k(\timo))(e^{\zeta \ti} - e^{\zeta \timo})\smin^2,
\end{equation}
where $\zeta = 2\ln(\frac{\smin}{\smax}) + 2\gamma_0$.

\section{Acknowledgements}
Funded by the Deutsche Forschungsgemeinschaft (DFG, German Research Foundation) 498394658, 545210893; by the Federal Ministry for Economic Affairs and Climate Action (Bundesministerium f\"ur Wirtschaft und Klimaschutz), Zentrales Innovationsprogramm Mittelstand (ZIM), Germany, within the project FKZ KK5528802VW4; and by the German Federal Ministry of Research, Technology and Space (BMFTR) under grant agreement No. 01IS24072A (COMFORT).

The authors gratefully acknowledge the scientific support and HPC resources provided by the Erlangen National High Performance Computing Center (NHR@FAU) of the Friedrich-Alexander-Universit\"at Erlangen-N\"urnberg (FAU) under the NHR projects f101ac. NHR funding is provided by federal and Bavarian state authorities. NHR@FAU hardware is partially funded by the German Research Foundation (DFG) - 440719683.

\section{Generative AI Use Disclosure}
Generative AI tools were used solely for language editing and polishing of the manuscript. These tools assisted in improving grammar, clarity, and readability. They were not used to generate scientific content, derive results, conduct experiments, or formulate the core contributions of this work. All authors take full responsibility for the content of this paper and have reviewed and approved the final manuscript.

\bibliographystyle{IEEEtran}
\bibliography{mybib}

\end{document}